\documentclass[letter,traditabstract,longauth]{aa} 
\usepackage{graphicx}
\usepackage{natbib}

\usepackage{txfonts}
\begin{document}
\newcommand{\Zsolar}{\mbox{$\,\rm Z_{\odot}$}}
\newcommand{\Msolar}{\mbox{$\,\rm M_{\odot}$}}
\newcommand{\Lsolar}{\mbox{$\,\rm L_{\odot}$}}
\newcommand{\xs}{$\chi^{2}$}
\newcommand{\dxs}{$\Delta\chi^{2}$}
\newcommand{\xsn}{$\chi^{2}_{\nu}$}
\newcommand{\ls}{{\tiny \( \stackrel{<}{\sim}\)}}
\newcommand{\gs}{{\tiny \( \stackrel{>}{\sim}\)}}
\newcommand{\asec}{$^{\prime\prime}$}
\newcommand{\amin}{$^{\prime}$}
\newcommand{\mstar}{\mbox{$M_{*}$}}
\newcommand{\hi}{H{\sc i}\ }
\newcommand{\hii}{H{\sc ii}\ }
\newcommand{\kms}{$\rm km~s^{-1}$}

   \title{The Herschel Virgo Cluster Survey: II. Truncated dust disks in H{\sc i}-deficient spirals\thanks{{\it Herschel} is an ESA space observatory with science 
   instruments provided by European-led Principal Investigator consortia and with important participation from NASA.}}

\author{
L. Cortese\inst{1}
\and
J. I. Davies\inst{1}
\and
M. Pohlen\inst{1}
\and
M. Baes\inst{2} 
\and
G. J. Bendo\inst{3}
\and
S. Bianchi\inst{4}
\and
A. Boselli\inst{5}
\and
I. De Looze\inst{2}
\and
J. Fritz \inst{2}
\and
J. Verstappen\inst{2}
\and
D. J. Bomans\inst{6}
\and
M. Clemens\inst{7}
\and
E. Corbelli\inst{4}
\and
A. Dariush\inst{1}
\and
S. di Serego Alighieri\inst{4}
\and
D. Fadda\inst{8}
\and
D. A. Garcia-Appadoo\inst{9}
\and
G. Gavazzi\inst{10}
\and
C. Giovanardi\inst{4}
\and
M. Grossi\inst{11}
\and
T. M. Hughes\inst{1}
\and
L. K. Hunt\inst{4}
\and
A. P. Jones\inst{12}
\and
S. Madden\inst{13}
\and
D. Pierini\inst{14}
\and
S. Sabatini\inst{15}
\and
M. W. L. Smith\inst{1}
\and
C. Vlahakis\inst{16} 
\and
E. M. Xilouris\inst{17}
\and
S. Zibetti\inst{18}
}

\institute{
School of Physics and Astronomy, Cardiff University, The Parade, Cardiff, CF24 3AA, UK
\and
Sterrenkundig Observatorium, Universiteit Gent, Krijgslaan 281 S9, B-9000 Gent, Belgium 
\and
Astrophysics Group, Imperial College London, Blackett Laboratory, Prince Consort Road, London SW7 2AZ, UK 
\and
INAF-Osservatorio Astrofisico di Arcetri, Largo Enrico Fermi 5, 50125 Firenze, Italy 
\and
Astronomical Institute, Ruhr-University Bochum, Universitaetsstr. 150, 44780 Bochum, Germany 
\and
Laboratoire d'Astrophysique de Marseille, UMR 6110 CNRS, 38 rue F. Joliot-Curie, F-13388 Marseille, France 
\and
INAF-Osservatorio Astronomico di Padova, Vicolo dell'Osservatorio 5, 35122 Padova, Italy
\and
NASA Herschel Science Center, California Institute of Technology, MS 100-22, Pasadena, CA 91125, USA 
\and
ESO, Alonso de Cordova 3107, Vitacura, Santiago, Chile 
\and
Universita' di Milano-Bicocca, piazza della Scienza 3, 20100, Milano, Italy 
\and
CAAUL, Observat\'orio Astron\'omico de Lisboa, Universidade de Lisboa, Tapada da Ajuda, 1349-018, Lisboa, Portugal
\and
Institut d'Astrophysique Spatiale (IAS), Batiment 121, Universite Paris-Sud 11 and CNRS, F-91405 Orsay, France 
\and
Laboratoire AIM, CEA/DSM- CNRS - Universit\'e Paris Diderot, Irfu/Service d'Astrophysique, 91191 Gif sur Yvette, France 
\and
Max-Planck-Institut fuer extraterrestrische Physik, Giessenbachstrasse, Postfach 1312, D-85741, Garching, Germany
\and
INAF-Istituto di Astrofisica Spaziale e Fisica Cosmica, via Fosso del Cavaliere 100, I-00133, Roma, Italy 
\and
Leiden Observatory, Leiden University, P.O. Box 9513, NL-2300 RA Leiden, The Netherlands 
\and
Institute of Astronomy and Astrophysics, National Observatory of Athens, I. Metaxa and Vas. Pavlou, P. Penteli, GR-15236 Athens, Greece 
\and
Max-Planck-Institut fuer Astronomie, Koenigstuhl 17, D-69117 Heidelberg,  Germany 
}

   \date{Submitted to A\&A Herschel Special Issue}

 
  \abstract{By combining {\it Herschel}-SPIRE observations obtained as part of the Herschel Virgo Cluster Survey 
  with 21 cm H{\sc i} data from the literature, we investigate the role of the cluster environment on the dust content of 
  Virgo spiral galaxies. We show for the first time that the extent of the dust disk is significantly reduced 
  in H{\sc i}-deficient galaxies, following remarkably well the observed `truncation' of the H{\sc i} disk. 
  The ratio of the submillimetre-to-optical diameter correlates with the H{\sc i}-deficiency, suggesting that 
  the cluster environment is able to strip dust as well as gas. 
These results provide important insights not only into the evolution of cluster galaxies but 
also into the metal enrichment of the intra-cluster medium.}


	\authorrunning{Cortese et al.}	
	\titlerunning{HeViCS II: Truncated dust disks in H{\sc i}-deficient spirals}
   \maketitle
%

\section{Introduction}
It is now well established that the evolution of 
spiral galaxies significantly depends on the environment they inhabit. 
The reduction in the star formation rate (e.g., \citealp{lewis02}) 
and atomic hydrogen (H{\sc i}) content  (e.g., \citealp{giova85}) of galaxies when moving 
from low- to high-density environments indicates that clusters are 
extremely hostile places for star-forming galaxies.
However, a detailed knowledge of the effects of the environment on all 
the components of the interstellar medium (ISM) is still lacking.
Particularly important is our understanding of how the environment is able 
to affect the dust content of cluster spirals. 
Dust plays an important role in the process of star formation, since it acts 
as a catalyzer for the formation of molecular hydrogen (H$_{2}$, from which stars are formed) 
and prevents its dissociation by the interstellar radiation field. Thus, the stripping 
of dust might significantly affect the properties of the ISM in infalling cluster spirals. 

Since dust is generally associated with the cold gas component of the ISM, it is 
expected that when the H{\sc i} is stripped part of the dust will be removed as well, but 
no definitive evidence of a reduced dust content in cluster galaxies has been found so far.
For a fixed morphological type, H{\sc i}-deficient galaxies\footnote{The 
H{\sc i}-deficiency ($def_{\rm HI}$) is defined as the difference, in 
logarithmic units, between the observed H{\sc i} mass and the value expected from an isolated galaxy 
with the same morphological type and optical diameter \citep{haynes}.} 
appear to have higher IRAS $f(100~\mu {\rm m})/f(60~\mu {\rm m})$ flux density ratios 
(i.e., colder dust temperatures, \citealp{bicay87}) and lower 
far-infrared (FIR) flux densities per unit optical area \citep{doyon89} than gas-rich galaxies.
However, by using ISO observations of the Virgo cluster \citep{tuffs02}, \cite{popescu02} find 
no strong variation with cluster-centric distance in the dust properties of each morphological type. 
Only the most extreme H{\sc i}-deficient galaxies appear to be lacking a cold dust component.
More recently, \cite{review} have revealed an interesting trend of decreasing 
dust masses per unit of H-band luminosity with decreasing distance from the center of Virgo.
Thus, it is still an open issue whether or not dust is removed from infalling cluster spirals.
 

The launch of {\it Herschel} \citep{pilbratt10} has opened a new era in the study of environmental 
effects on dust. 
Thanks to its high spatial resolution and sensitivity to all dust components,
{\it Herschel} will be able to determine if cluster galaxies have lost a significant amount of their dust content. 
Ideally, this analysis should be done on a large, statistically complete sample, following the same 
criteria used to define the H{\sc i}-deficiency parameter \citep{haynes}: i.e., by comparing the 
dust content of galaxies of the same morphological type but in different environments.
By observing a significant fraction ($\sim$64 deg$^2$) of the Virgo cluster 
at 100, 160, 250, 350 and 500 $\mu$m, the Herschel Virgo Cluster Survey (HeViCS, \citealp{davies10}, hereafter Paper I; 
see also http://www.hevics.org) 
will soon provide the optimal sample for such an investigation. 
In the meantime, with the first HeViCS data it is possible to use a more indirect approach and 
compare the extent of the dust disk in gas-rich and gas-poor cluster galaxies.  
Since previous studies have shown that the H{\sc i} stripping is associated with a `truncation'\footnote{The term `truncation' 
is used here to indicate either an abrupt steepening of the surface-brightness profile or, more simply, a 
significant reduction in the disk scale-length compared to the optical one.} of 
the gas \citep{cayatte94} and star-forming disk \citep{koopmann04,cati05,review}, if the dust follows 
the atomic hydrogen we should find a reduction in the extent of the dust disk with increasing H{\sc i}-deficiency.
\begin{figure*}
  \centering
  \includegraphics[width=17cm]{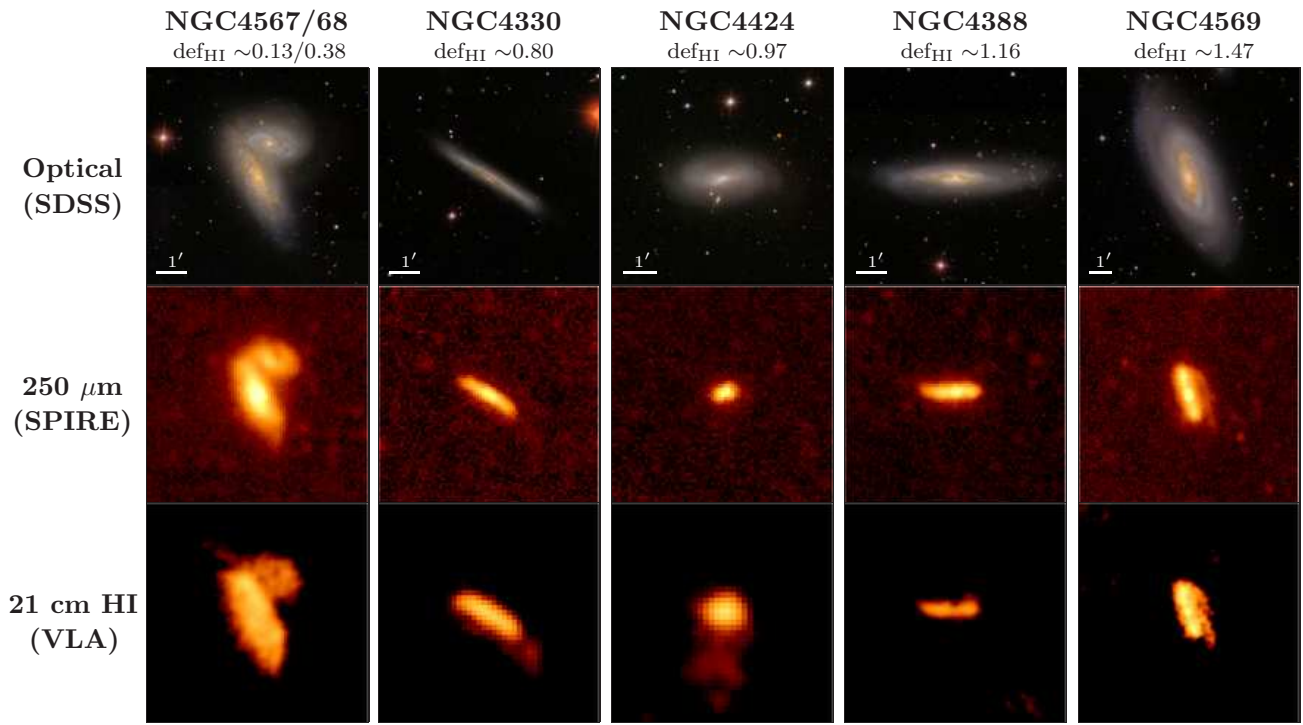}
     \caption{The optical (top), 250 $\mu$m (middle) and H{\sc i} (bottom) maps for galaxies in our sample with different degrees of H{\sc i}-deficiency.}
	 \label{fc}
  \end{figure*}

In this paper we will take advantage of the HeViCS observations, obtained as 
part of the {\it Herschel} Science Demonstration (SD) phase, 
to investigate the correlation between the dust distribution and gas content in cluster galaxies.


\section{Observations and data reduction}
A $\sim$245\arcmin x 230\arcmin~field in the center of the Virgo cluster has been observed 
by {\it Herschel} using the SPIRE/PACS \citep{griffin10,poglitsch10} parallel scan-map mode 
as part of the SD observations for HeViCS.
In this paper we will focus our attention on the 3 SPIRE bands only.
The full widths at half maximum of the SPIRE beams are 18.1, 25.2, 36.9 
arcsec at 250, 350 and 500 $\mu$m, respectively.
Details about the observations and data reduction can be found in Paper I.
The typical rms noise across the whole image 
are $\sim$12, 10, 12 mJy/beam at 250, 350 and 500 $\mu$m, respectively (i.e., $\sim$2 times 
higher than the confusion noise).
No spatial filtering is applied during the data reduction, making SPIRE 
maps ideal to investigate extended submillimetre (submm) emission.
The uncertainty in the flux calibration is of the order of 15\%.

In order to investigate how the dust distribution varies with the degree of H{\sc i}-deficiency 
in Virgo spirals, we restricted our analysis to the 15 spiral galaxies in the 
HeViCS SD field for which H{\sc i} surface density profiles are available.
The H{\sc i} maps are obtained from the recent `VLA Imaging of Virgo in Atomic gas' (VIVA) 
survey (\citealp{chung09}, 13 galaxies: NGC4294, NGC4299, NGC4330, NGC4351, NGC4380, NGC4388, NGC4402, 
NGC4424, NGC4501, NGC4567, NGC4568, NGC4569, NGC4579), from \citet[NGC4438]{cayatte94} and from \citet[NGC4413]{warmels88}.
H{\sc i}-deficiencies have been determined following the prescription presented in \cite{chung09}. 
This method is slightly different from the original definition presented by \cite{haynes}, as it assumes 
a mean H{\sc i} mass-diameter relation, regardless of the morphological type. 
Following \cite{chung09}, we use the difference between 
the type-dependent and type-independent definitions as uncertainty in the H{\sc i}-deficiency parameter. 
We note that, on average, this value is smaller 
than the intrinsic scatter of $def_{\rm HI}$ for field galaxies ($\sim$0.27, \citealp{fumagalli09}).

Surface brightness profiles in the three SPIRE bands were derived using the IRAF task \textsc{ellipse}. 
The center was fixed to the galaxy's optical center (taken from the NASA/IPAC Extragalactic Database\footnote{http://nedwww.ipac.caltech.edu/}) and 
the ellipticity and 
position angle to the same values adopted for the H{\sc i} profiles taken from the literature \citep{chung09,cayatte94,warmels88}.
The sky background was determined within rectangular regions around the galaxy and subtracted from 
the images before performing the ellipse fitting. Each profile was then corrected to the `face-on' value using 
the inclinations taken from the literature.
All the galaxies in our sample are clearly resolved in all the three SPIRE bands: e.g., 
on average $\sim$4-5 beam sizes at 500 $\mu$m.
Submm isophotal radii were determined at 6.7$\times$10$^{-5}$, 3.4$\times$10$^{-5}$ 
and 1.7$\times$10$^{-5}$ Jy arcsec$^{-2}$ surface brightness level in 250, 350 and 500 $\mu$m respectively.
These are the average surface brightnesses observed at the optical radius 
(25 mag arcsec$^{-2}$ in B band, \citealp{rc3}) in the four non H{\sc i}-deficient 
galaxies ($def_{\rm HI}<$0.3) in our sample (NGC4294, 4299, 4351, 4567) and 
correspond to $\sim$2-3$\sigma$ noise level.   
Of course, this choice is rather arbitrary and it has 
no real physical basis. However, as discussed in \S~3, the result does not 
depend on the way in which the isophotal radii have been defined.
Although many of our galaxies show some evidence of nuclear activity, we do not find a single 
case in which the nuclear submm emission dominates the emission from the disk (see also \citealp{sauvage10}). 
Thus, the isophotal radius 
is a fair indication of the extent of the dust disk.

\section{Results \& discussion}
\begin{figure}
  \centering
  \includegraphics[width=7.25cm]{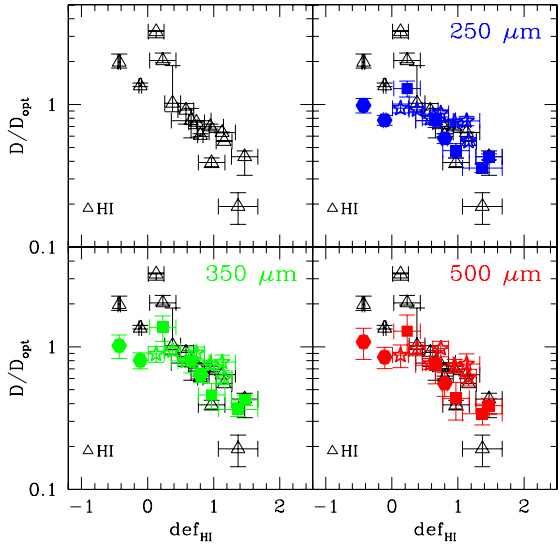}
     \caption{The ratio of the submm-to-optical diameters versus H{\sc i}-deficiency in the three SPIRE 
     bands. Squares are for Sa-Sab, stars for Sb-Sbc and hexagons for Sc and later types. 
     For comparison, the triangles show the same relation for  
     the H{\sc i}-to-optical diameter ratio, where the H{\sc i} isophotal diameters 
     are taken at a surface density level of 1 M$_{\odot}$ pc$^{-2}$ \citep{chung09}.}
	 \label{isoradius}
  \end{figure}

\begin{figure}
  \centering
  \includegraphics[width=8cm]{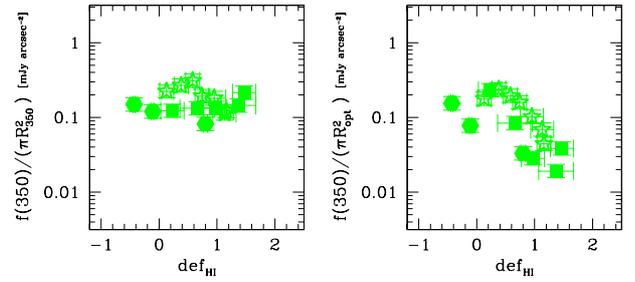}
     \caption{The 350 $\mu$m flux per unit of 350 $\mu$m area (left) and optical area (right) versus H{\sc i}-deficiency. Symbols are as in
     Fig.~\ref{isoradius}.}
	\label{SB}
  \end{figure}

\begin{figure}
  \centering
  \includegraphics[width=7.6cm]{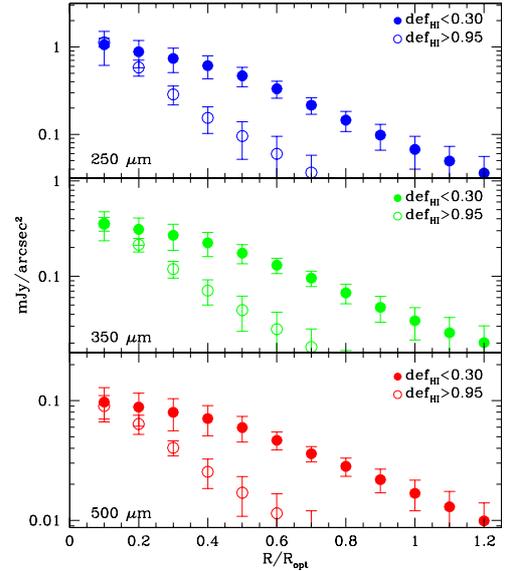}
     \caption{Average submm surface-brightness profiles in bins of normalized radius for normal and highly H{\sc i}-deficient 
     galaxies.}
	\label{profiles}
  \end{figure}

In Fig.~\ref{fc} we compare the optical, 250 $\mu$m and H{\sc i} maps for  
a subsample of our galaxies with different levels of H{\sc i}-deficiency. In 
highly deficient spirals the 250 $\mu$m emission is significantly less extended than the optical, 
following remarkably well the observed `truncation' of the H{\sc i} disk\footnote{See also \cite{pohlen10} for 
an analysis of the two grand design Virgo spirals NGC4254 ($def_{\rm HI}\sim-$0.10) and NGC4321 ($def_{\rm HI}\sim$0.35).}. 
This is confirmed in Fig.~\ref{isoradius}, where we show the ratio of the 
submm-to-optical isophotal diameters as a function of $def_{HI}$ for the 15 galaxies in our sample. 
For all the three SPIRE bands we find a strong correlation (Spearman correlation coefficient $r_{s}\sim-$0.87, 
corresponding to a probability $P(r>r_{s})>$ 99.9\% that the two variables are correlated) between 
the submm-to-optical diameter ratio and $def_{HI}$.
Although qualitatively supported by Fig.~\ref{fc}, this correlation alone does not imply a change in the shape of the submm profile. 
A decrease in the central submm surface brightness of gas-poor galaxies could produce a similar trend without the need to invoke a  
reduction in the disk scale-length. However, Fig.~\ref{SB} and \ref{profiles} clearly exclude such a scenario.
In Fig.~\ref{SB} we show that, while the 350 $\mu$m flux per unit of 350 $\mu$m area (i.e., the average submm surface brightness) 
is nearly constant across the whole sample, the 350 $\mu$m flux per unit of optical area significantly decreases with 
increasing $def_{HI}$.
This is even more evident in Fig.~\ref{profiles} where the average surface brightness profiles in bins of normalized 
radius for gas-rich and gas-poor galaxies ($def_{HI}>$0.96; i.e., NGC4380, NGC4388, NGC4424, NGC4438, NGC4569) are shown. 
While the central surface brightness is approximately the same, the profile of H{\sc i}-deficient 
galaxies is steeper than in normal galaxies and falls below our detection limit at approximately half 
the optical radius\footnote{This also confirms that the correlation shown in Fig.~\ref{isoradius} is not qualitatively 
affected by the definition of submm isophotal radius adopted here.}.
We can thus conclude that H{\sc i}-deficient galaxies have submm disks significantly less extended than the 
optical disks, following closely the `truncation' observed in H{\sc i}.

Interestingly, from Fig.~\ref{isoradius} and \ref{SB} it emerges that the extent of the dust disk is significantly reduced 
compared to the optical disk only for high H{\sc i}-deficiencies ($def_{\rm HI}$\gs0.8-1), i. e.
when the atomic hydrogen starts to be stripped from inside the optical radius.

We now need to consider whether we are just observing a trend due to a different mix of morphologies between gas-rich and 
gas-poor galaxies. 
Although H{\sc i}-deficient systems are of earlier type than gas-rich spirals, our result does not change if we focus our 
attention on Sa-Sbc galaxies only (i.e., 80\% of our sample). 
Since in this range the average 850 $\mu$m scale-length-to-optical radius \citep{thomas04} and H{\sc i}-to-optical 
radius \citep{cayatte94} ratios do not vary significantly (i.e., less than 1$\sigma$) with galaxy type, morphology 
alone cannot be responsible for the correlations shown in Fig.~\ref{isoradius} and \ref{SB}. 
Moreover, all the highly H{\sc i}-deficient galaxies in our sample are well known perturbed Virgo spirals, 
on which the influence of the cluster environment has already been proven (e.g., \citealp{vollmer09}).  
So, the difference in the dust distribution between gas-poor and gas-rich spirals observed here 
is likely due to the effect of the cluster environment and is not just related to the intrinsic 
properties of each galaxy. Future analysis of a larger and more complete sample 
will allow us to further disentangle the role of environment from morphology on the dust 
distribution in nearby spirals.

A `truncation' in the surface brightness profile of NGC4569 (the most H{\sc i}-deficient galaxy 
in our sample) has already been observed at {\it Spitzer} 24 and 70 $\mu$m by \cite{n4569}. 
However, while a reduction in the 24 and 70 $\mu$m surface brightness 
might just be a direct consequence of the quenching of the star formation in gas-poor 
galaxies, this scenario is not valid in our case.
For $\lambda$\gs100-200 $\mu$m, the dust emission does not come predominantly 
from grains directly heated by photons associated with star formation activity, but from a 
colder component heated by photons of the diffuse interstellar radiation field  
(e.g., \citealp{chini86,draine07,bendo10}). 
Since this colder component dominates the dust mass budget in galaxies, 
the trends here observed are likely not due to a 
reduction in the intensity of the ultraviolet radiation field, but they imply that in 
H{\sc i}-deficient galaxies the dust surface density in the outer parts of the disk 
is significantly lower than in normal spirals.

\begin{figure}
  \centering
  \includegraphics[width=8cm]{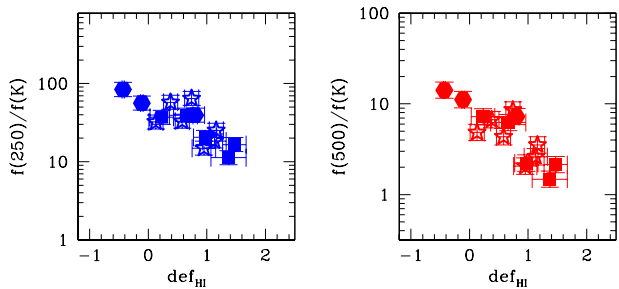}
     \caption{The 250 $\mu$m-to-K-band (left) and 500 $\mu$m-to-K-band flux density ratios versus H{\sc i}-deficiency. 
     Symbols are as in Fig.~\ref{isoradius}.}
	\label{colours}
  \end{figure}

An alternative way to compare the properties of normal and gas-poor Virgo spirals is 
to look at their submm-to-near-infrared colours. 
Since the K-band is an ideal proxy for the stellar mass and the SPIRE fluxes provide an indication 
of the total dust mass, it is interesting to investigate how the $f(250)/f(K)$ and 
$f(500)/f(K)$ flux density ratios vary with $def_{HI}$. 
We find that highly H{\sc i}-deficient galaxies have $f(250)/f(K)$ and $f(500)/f(K)$ ratios a 
factor $\sim$2-3 lower than normal galaxies (Fig.~\ref{colours}). 
This provides additional support to a scenario in which gas-poor galaxies have also lost 
a significant fraction of their original dust content. 

By comparing the dust mass per unit of H-band luminosity for a sample of late-type 
galaxies in the Coma-Abell1367 supercluster, \cite{contursi01} find no significant difference 
in the dust content of normal and H{\sc i}-deficient spirals, apparently in contrast 
with our results. However, such a difference is due (at least in part) to the 
fact that the sample used by \cite{contursi01} does not include any galaxy with $def_{\rm HI}>$0.87.
It is easy to see in Fig.~\ref{colours} that, for $def_{\rm HI}<$0.87, almost no trend is observed between 
the $f(500)/f(K)$ (or $f(250)/f(K)$) flux density ratio and $def_{HI}$. 
In fact, the Spearman correlation coefficient drops from  $r_{s}\sim-$0.78 to $\sim-$0.28 
and $\sim-$0.11 (i.e., a drop in the probability that two variables are correlated to 80 and 40\%) 
for the $f(250)/f(K)$ and $f(500)/f(K)$ ratios, respectively.
This implies that the two variables are no longer significantly correlated, highlighting 
once more that substantial dust stripping is observed only if the ISM is removed from well 
within the optical radius.

\section{Conclusions}
In this paper, we have shown that in H{\sc i}-deficient galaxies the dust disk is significantly less 
extended than in gas-rich systems. This result, combined with the evidence that H{\sc i}-deficient objects 
show a reduction in their submm-to-K-band flux density ratios, suggests that when 
the atomic hydrogen is stripped part of the dust is removed as well. 
However, the dust stripping appears efficient only when very gas-poor spirals are considered, 
implying that in order to be significant the stripping has to occur well within the optical radius. 
This is consistent with \cite{thomas04} who found that the 850 $\mu$m scale-length of nearby 
galaxies is smaller than the H{\sc i}, suggesting that outside the optical 
radius the gas-to-dust ratio is higher than in the inner parts.

Our analysis provides evidence that the cluster environment is able to significantly 
alter the dust properties of infalling spirals. We note that this has only been possible thanks 
to the unique spatial resolution and high sensitivity in detecting cold dust provided by the {\it Herschel}-SPIRE 
instrument and to the wide range of H{\sc i}-deficiencies covered by our sample.
Once combined with the direct detection of stripped dust presented by \cite{cortese10b} and 
\cite{gomez10}, our results highlight dust stripping by environmental effects 
as an important mechanism for injecting dust grains into the intra-cluster medium, 
thus contributing to its metal enrichment. This is consistent with numerical simulations 
which predict that ram pressure alone can already contribute $\sim$10\% of the enrichment 
of the ICM in clusters \citep{domainko}.
Interestingly, the stripped grains should survive in the hot ICM long 
enough to be observed \citep{popescu00,clemens10}.
 
Once completed, HeViCS will allow a search for additional evidence 
of dust stripping and place important constraints on the amount of intra-cluster dust present in Virgo. 
Moreover, in combination with the Herschel Reference Survey \citep{hrs}, 
it will be eventually possible to accurately quantify the degree of dust-deficiency in Virgo spirals. 

\begin{acknowledgements}
We thank the referee, Richard Tuffs, for useful comments which improved the 
clarity of this manuscript.
We thank all the people involved in the construction and launch of {\it Herschel}. 
In particular, the {\it Herschel} Project Scientist G. Pilbratt, and the PACS and SPIRE instrument 
teams.
\end{acknowledgements}

\end{document}